\documentstyle[epsfig]{aipproc}

\begin{document}


\title{Spatially Varying X-ray Synchrotron Emission in SN~1006}

\author{Kristy K. Dyer, Stephen P. Reynolds, Kazik J. Borkowski$^1$ and Robert Petre$^2$}
\address{$^1$ North Carolina State University, Physics Department Box 8202, Raleigh NC 27695 \\
$^2$ NASA's GSFC, LHEA Code 666, Greenbelt MD 20771}

\lefthead{To appear in ``Young Supernova Remnants'' }
\righthead{11th Oct. Maryland AIP ed. S. Holt \& U. Hwang}


\maketitle

\begin{abstract}

A growing number of both galactic and extragalactic supernova remnants show non-thermal (non-plerionic) emission in the X-ray band. New synchrotron models, realized as {\it SRESC} and {\it SRCUT} in XSPEC 11, which use the radio spectral index and flux as inputs and include the full single-particle emissivity, have demonstrated that synchrotron emission is capable of
producing the spectra of dominantly non-thermal supernova remnants with
interesting consequences for residual thermal abundances and acceleration
of particles. In addition, these models deliver a much better-constrained separation between the thermal and non-thermal components, whereas combining an
unconstrained powerlaw with modern thermal models can produce a range of
acceptable fits.  While synchrotron emission can be approximated by a powerlaw over small ranges of energy, the synchrotron spectrum is in fact steepening over the X-ray band. Having demonstrated that the integrated spectrum of
SN~1006, a remnant dominated by non-thermal emission, is well described by
synchrotron models I now turn to spatially resolved observations of
this well studied remnant. The synchrotron models make both spectral and
spatial predictions, describing how the non-thermal emission varies across
the remnant. Armed with spatially resolved non-thermal models and new
thermal models such as {\it VPSHOCK} we can now dissect the inner workings of
SN~1006.

\end{abstract}

\section*{Versions of {\it SRESC} for subregions of SN~1006}

In Dyer et al. 2000\cite{Dyer} we used new X-ray synchrotron models along with information from the radio synchrotron spectrum to determine with a new degree of accuracy the total amount of synchrotron emission in the spectrum of SN~1006. This analysis can be extended to regions of the supernova remnant (SNR). Here we separate the SNR into five regions -- north and south limbs (northeast and southwest), north and south polar caps and center, as shown in Figure 1. Chandra observations of SNRs will be able to analyze much smaller regions, limited only by signal to noise, so these methods of spatial-spectral analysis of complex thermal and nonthermal emission, may be necessary for other SNRs.




\begin{figure} 
\centerline{\qquad{\epsfig{file=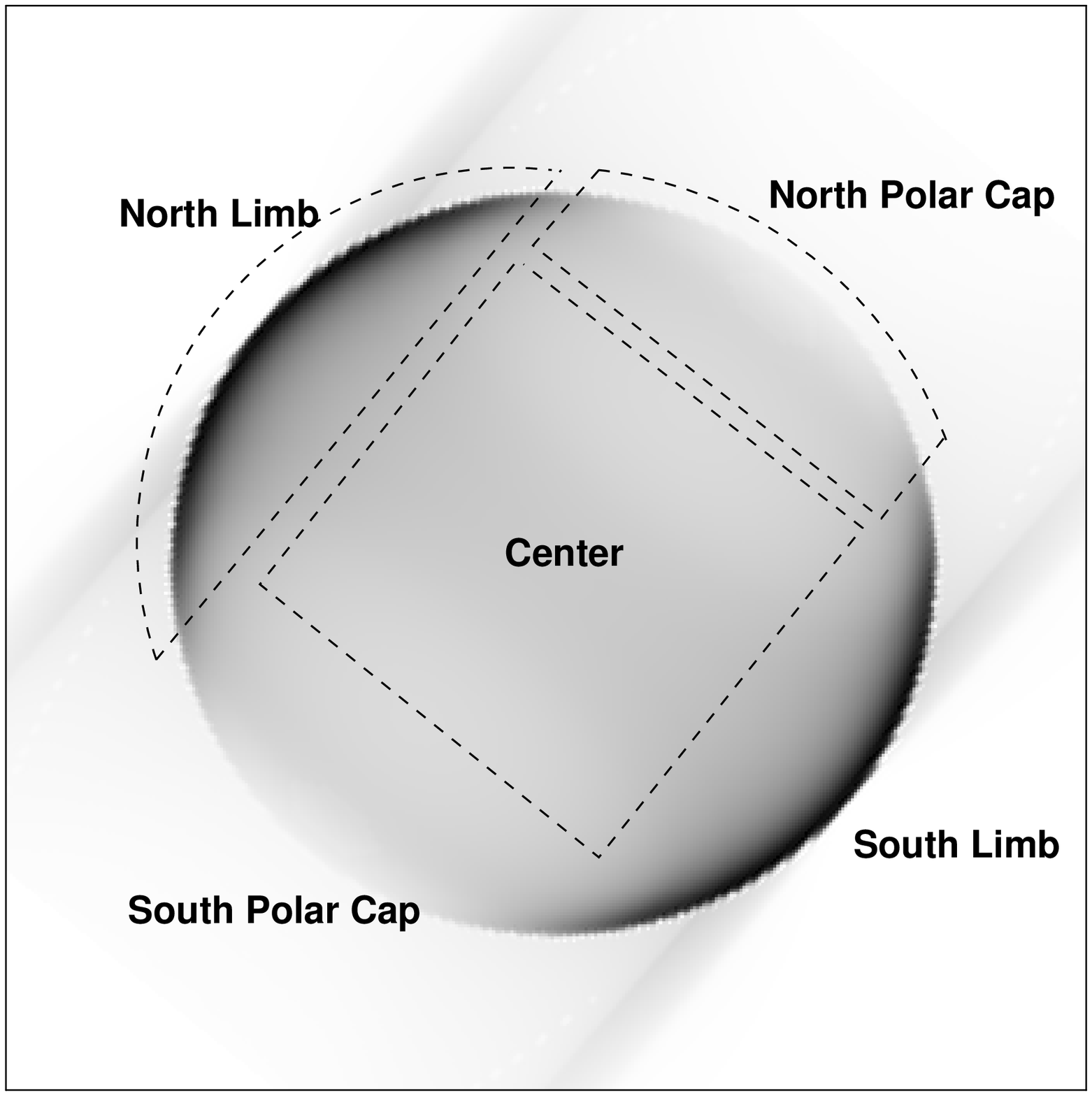,height=2.5in,width=2.5in}}
\quad{\epsfig{file=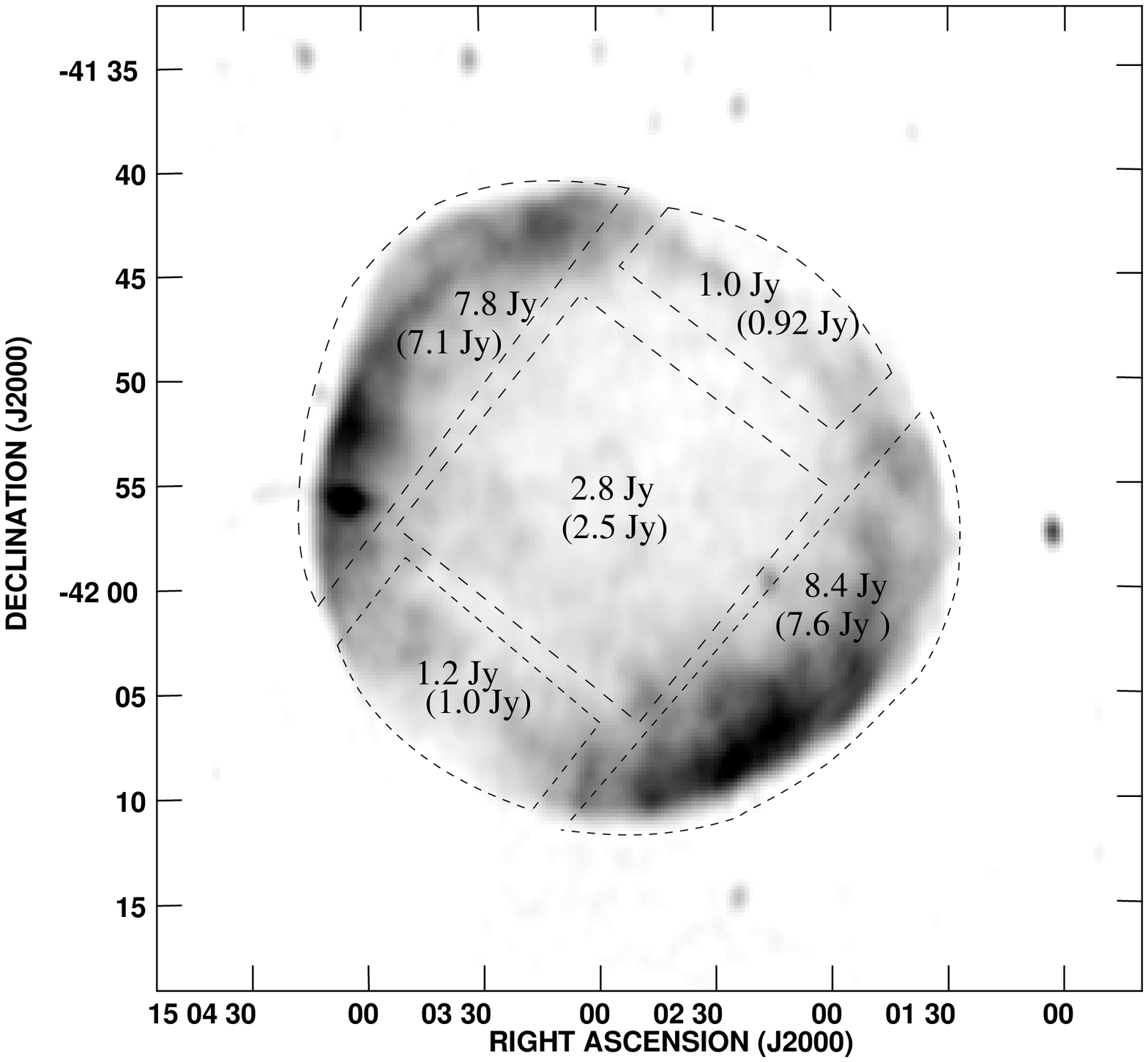,height=2.5in,width=2.8 in}}}
\vspace{10pt}
\caption{Left: Synchrotron emission for the simulated SNR.
Right: The Molonglo + Parkes radio image with the subregions indicated. The value shown is the measured flux at 843 MHz, in parenthesis is the scaled 1~GHz value used in the model. Thanks to R.S. Roger for recovering the SN~1006 map. }
\end{figure}




The synchrotron models were developed \cite{Reynolds1,Reynolds2} to describe the two dimensional distribution of synchrotron emission observed for a spherically symmetric SNR expanding in a uniform upstream magnetic field. The one-dimensional spectrum extracted for {\it SRESC} was was created by summing over the simulated SNR. Here we have developed subsets of the {\it SRESC} model: {\it LIMB}, {\it CENTER} and {\it CAP} (shown in Figure 1)  by summing over discrete regions in the simulated SNR. The synchrotron emission from each region has a slightly different spectrum since it samples different populations of electrons interacting with the upstream magnetic field at different obliquity angles.



As in the {\it SRESC} model, these synchrotron models define the nonthermal \mbox{X-ray} emission with three parameters: 
the radio spectral index $\alpha$, 1~GHz radio flux [Jy], and the rolloff [Hz].
The rolloff determines the curvature by specifying the frequency at which the model has dropped by a factor of $\sim$6 from a powerlaw.  

Many papers have investigated small changes in spectral index in radio synchrotron emission in SNRs\cite{Anderson}. However, flux uncertainties in interferometric measurements and nonlinear deconvolution processes have made unambiguous detections of these index changes difficult\cite{Dyer2}. In any case, the expected spectral index variation ($\Delta \alpha \pm 0.01$) are expected to be small compared to the uncertainties in the spectral index ($\alpha \pm 0.2$). Here we assume the spectral index is constant across the SNR, a prediction that does agree with shock acceleration theory.  
For integrated spectral analysis the 1~GHz~flux is readily available
\footnote{\mbox{Available for Galactic SNRs at Green's Catalogue} http://www.mrao.cam.ac.uk/surveys/snrs/index.html and for some Magellanic Cloud SNRs at http://www.renaissoft.com/snrc/}
. However, for spatially resolved spectral analysis the flux must be measured in each subregion. For large remnants such as SN~1006 this can be a problem. Maps from interferometers (such as the VLA) are unsuitable for measuring small scale fluxes, since interferometers are not sensitive to flux on angular scales beyond those corresponding to the smallest spacing. Correcting the image by adding a constant to bring it up to the single dish flux has not proven sufficiently accurate in small regions. The solution is to use images created with both interferometric and single dish data, which sample the flux at all spatial scales. Here we use a map made with the Molonglo interferometer plus the Parkes single dish telescope\cite{Roger} to measure the flux in subregions. The curvature of the X-ray synchrotron is determined by {\it rolloff}. This factor was fixed by fitting a small region in the north limb expected to be dominated by synchrotron emission, and we have assumed, without evidence or theoretical expectation to the contrary, that this value is constant across the SNR.  

\section*{Results}

These models reveal that there is significant nonthermal emission throughout SN~1006. Even the polar caps and center, which have prominent thermal lines, contain a synchrotron component (see Table \ref{table1}). As might be expected, it is most difficult to determine the amount of synchrotron emission in regions where thermal emission dominates. Model fits of the center of SN~1006 will tolerate up to 16\% of the flux in synchrotron. Differences in temperature and abundances between the fits may make it possible to argue for or against a synchrotron component for the center of the remnant. 

\begin{table}
\caption{Predicted nonthermal X-rays in regions of SN~1006}\label{table1}
\begin{tabular}{lcc}
   Region&
   \multicolumn{1}{c}{ sub-models of {\it SRESC}\tablenote{For $\alpha$=0.60 and rolloff (measured from the north limb) of 3$\times 10^{17}$ Hz. See text for discussion.}} &
  \multicolumn{1}{c}{Nonthermal Flux (\% of total)}\\
\tableline
Center		& {\it CENTER}	&  $<$16 \cr
North Polar Cap	& {\it CAP}	&  13 \cr
South Polar Cap	& {\it CAP}	&  8 \cr
North Limb	& {\it LIMB}	&  37 \cr
\end{tabular}
\end{table}


Unfortunately there are no entirely satisfactory thermal models available to describe emission from Type Ia supernova remnants, such as SN~1006.  Any attempts to model the thermal emission run into several complications: Young SNRs such as SN~1006 may require solutions intermediate between self-similar driven wave and Sedov\cite{Dwarkadas}. Integrated fits to the full remnant showed high abundances \cite{Dyer} suggesting a reverse shock, so a two-shock description may be necessary.  In light of this we use the plane shock model {\it VPSHOCK} as a starting point, making note of the fact  it allows a range of ionization timescales, and can describe heavy element ejecta, but assumes a constant density and temperature. 

\begin{figure} 
\centerline{{\epsfig{file=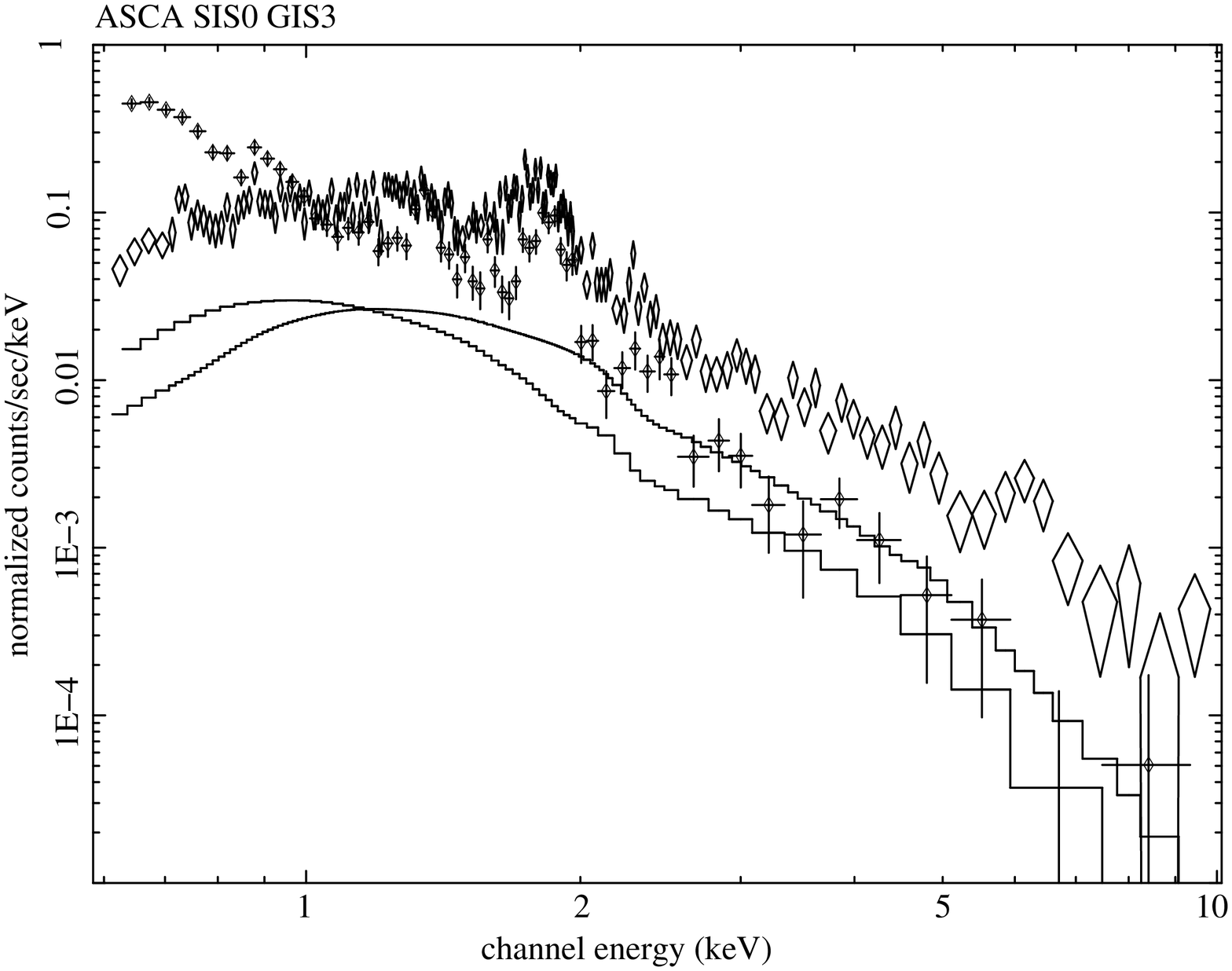,height=2.2in,width=3.0in}}
\quad{\epsfig{file=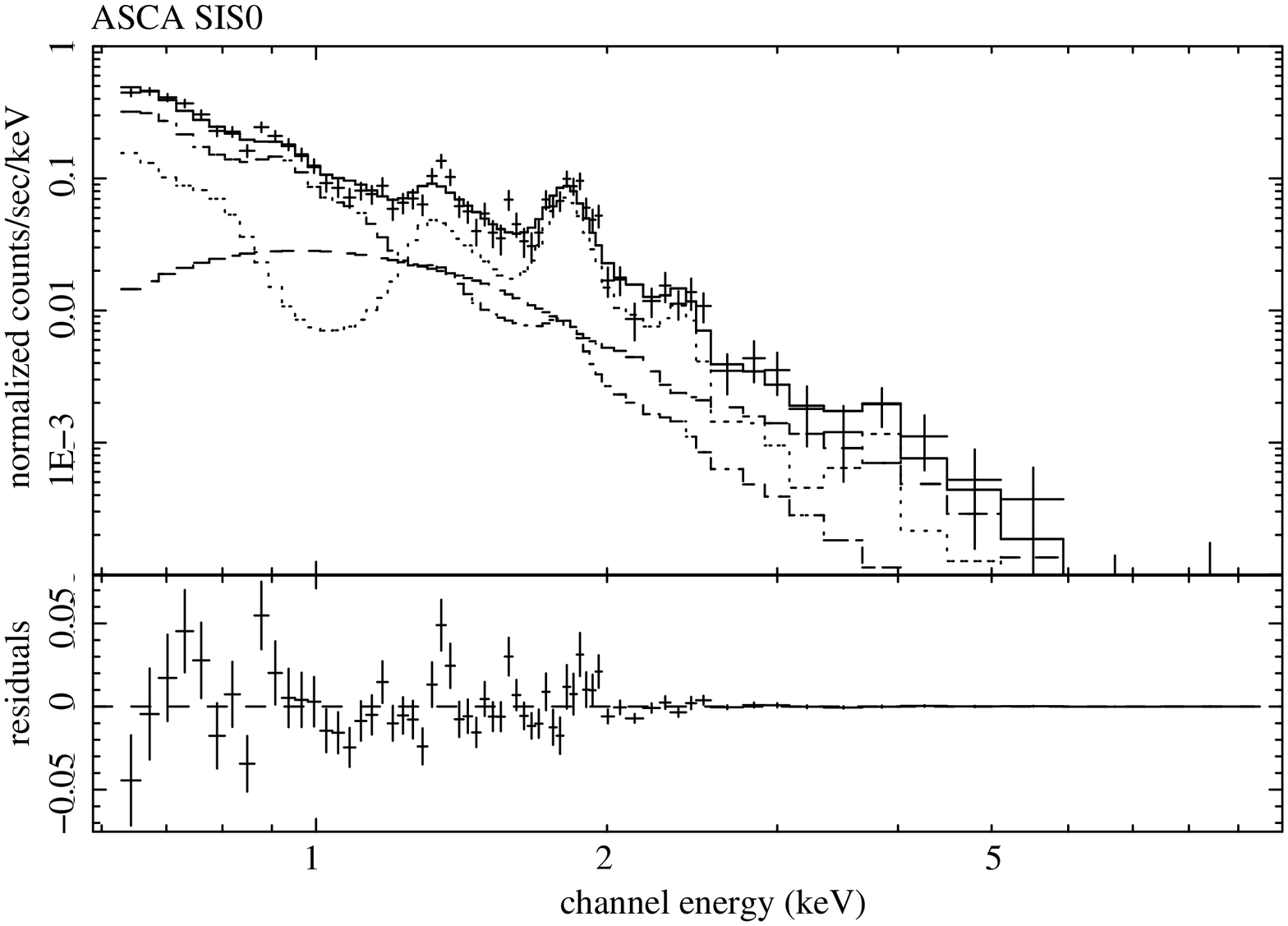,height=2.2in,width=2.5 in}}}
\vspace{10pt}
\caption{ASCA spectra of the South Polar Cap region of SN~1006. Left: GIS and SIS spectrum of regions of SN~1006 shown with the model predicted non-thermal emission
Right: ASCA SIS  with the {\it CAP} synchrotron model plus two plane parallel shocks ({\it VPSHOCK}), a low temperature with high abundances and a solar abundances with a higher temperature.}
\end{figure}

\section*{Conclusion}


%

We have found that there is significant nonthermal emission in all regions of the SN~1006, not only the limbs. With these new models it is now possible to separate the thermal and nonthermal emission in spatially resolved regions (just in time for Chandra observations). This is an absolute prerequisite if the thermal emission is to be understood. This work allows us to address new questions, among them: TeV $\gamma$-rays have been observed from the north limb but not the south limb of SN~1006. Are there differences in the synchrotron emission from the two limbs? Preliminary results indicate that the curvature is different for synchrotron in the north and south limb. H$\alpha$ is observed at the edge of the north polar cap region. Does the X-ray description of the thermal shock agree with optical observations? 





\begin{references}

\bibitem{Dyer} Dyer, K.K., Reynolds, S.P., Borkowski, K.J., Petre, R., Allen, G.E. 2001 {\it Separating Thermal and Nonthermal X-Rays in Supernova Remnants I:  Total Fits to SN~1006 AD} {\bf astro-ph/0010424} Accepted for the Astrophysical Journal March 10, 2001 

\bibitem{Reynolds1} Reynolds, S.\ P.\ 1996,
ApJL, 459, L13  

\bibitem{Reynolds2} Reynolds, S.\ P.\ 1998, ApJ,
493, 375 

\bibitem{Anderson} Anderson. M. C., \& Rudnick, L. 1993, ApJ, 408, 514  


\bibitem{Dyer2} Dyer, K.\ K.\ \& Reynolds, S.\ P.\ 1999, ApJ, 526, 365

\bibitem{Roger} Roger, R.\ S., Milne, D.\ 
K., Kesteven, M.\ J., Wellington, K.\ J.\ \& Haynes, R.\ F.\ 1988, ApJ, 
332, 940 

\bibitem{Dwarkadas} Dwarkadas, V.\ V.\ \& Chevalier, R.\ A.\ 1998, ApJ, 497, 807 







\end{references}
\end{document}